\documentclass{JHEP3}
\newcommand{\beq}{\begin{equation}}
\newcommand{\eeq}{\end{equation}}
\newcommand{\phib}{\ensuremath{\overline{\phi}}}
\newcommand{\etab}{\ensuremath{\overline{\eta}}}
\newcommand{\psib}{\ensuremath{\overline{\psi}}}
\newcommand{\chib}{\ensuremath{\overline{\chi}}}
\newcommand{\KD}{K\"{a}hler-Dirac }

\usepackage{epsfig,multicol}

\title{Simulations of ${\cal N}=2$ super Yang-Mills theory in two dimensions}

\author{Simon Catterall\\
        Department of Physics, Syracuse University, Syracuse, NY 13244, USA\\
        E-mail: \email{smc@physics.syr.edu}\\
        }

\preprint{SU-4252-822}  

\abstract{We present results from lattice simulations of ${\cal N}=2$
super Yang-Mills theory in two dimensions. The lattice formulation we
use was developed in \cite{2dpaper} and retains both
gauge invariance and an exact (twisted)
supersymmetry for any lattice spacing. Results for both $U(2)$ and $SU(2)$
gauge groups are given. We focus on supersymmetric Ward identities,
the phase of the Pfaffian resulting from integration
over the Grassmann fields and the nature of
the quantum moduli space.
}


\begin{document} 
\section{Introduction}
Supersymmetric field theories play a central role in modern theories
of particle physics. They offer a possible solution to
the gauge hierarchy problem and are often
more tractable analytically
than their non-supersymmetric counterparts while still exhibiting
features like confinement and chiral symmetry breaking \cite{prop}.
Super Yang-Mills theories are especially interesting because of their
possible connection to string and M-theory \cite{M}. One example of this is
the conjectured equivalence between $(p+1)$-dimensional super Yang-Mills
theory and supergravity containing $Dp$-brane sources \cite{dual}.

One attractive scenario for embedding the Standard Model in such a theory
is to imagine that supersymmetry breaks spontaneously as some gauge
coupling becomes large at low energy. Unfortunately various non-renormalization
theorems forbid such a spontaneous breaking at any finite order of
perturbation theory so one must turn to non-perturbative mechanisms to
drive such symmetry breaking.

Of course the lattice furnishes perhaps the only generally applicable
way to
study non-perturbative dynamics in field theory. 
However, 
the difficulties of discretizing supersymmetric theories are well known.
Generic naive discretizations of continuum supersymmetric theories
do not preserve supersymmetry. Typically quantum corrections 
then generate a large
number of relevant supersymmetry violating interactions whose
couplings must be tuned to zero as the lattice spacing is reduced.
This is both unnatural and in many cases (especially for
models with extended supersymmetry) prohibitively difficult.
Various attempts have been made over the last twenty five years to
overcome these problems see \cite{old} and \cite{new} and the recent
reviews \cite{feo_rev,kap_rev,cat_rev}.

Quite recently a series of new approaches have been developed which share
the common feature of preserving a sub-algebra of the full supersymmetry
algebra {\it exactly} at finite lattice spacing\footnote{Very recently a
lattice construction of ${\cal N}=2$ super Yang-Mills in $D=2$ 
has been proposed which claims to
preserve {\it all} the supercharges \cite{n=2_kawa}}
In the approach pioneered by Kaplan and collaborators \cite{kap}
the lattice theory is derived by applying a carefully
chosen orbifold condition to a supersymmetric matrix model (see
also \cite{joeldet,joelsim,pert,mithat}). 
A second approach proceeds from a 
reformulation of the theory using ideas drawn from
topological field theory \cite{top}.\footnote{It appears that these
two approaches may be intimately connected -- 
private communication Mithat Unsal}. It is this approach that
is the focus of this paper. 
The hope is that the residual exact supersymmetry will protect
the theory from dangerous radiative corrections and obviate
the need for fine tuning \cite{kap,sugino,joelqm,joelwz}.

This technique was initially used for theories without gauge symmetry
\cite{qm,wz2,sigma}
corresponding to supersymmetric quantum mechanics, the 2D complex Wess-Zumino
model and supersymmetric sigma models. An implementation for gauge theories 
was initially given by Sugino \cite{sugino}. 
Important progress was made when Kawamoto and collaborators
\cite{twist_kawa} pointed out the connection between these topological
formulations and \KD fermions. This has led to
additional lattice formulations which emphasize the
geometrical nature of the underlying theory \cite{n=2_kawa} and
\cite{2dpaper,4dpaper} (Hamiltonian formulations of lattice supersymmetric
theories employing \KD fermions were first proposed in \cite{schwim} and
later in \cite{aratyn})

The key starting point in this approach is to construct a new rotation group
from a combination of the original rotation group and part of the
R-symmetry associated with the extended SUSY. The supersymmetric field
theory is then reformulated in terms of fields which transform as
integer spin representations of this new rotation group
\cite{witten,conf,twist_kawa}.
This process is given the name {\it twisting} and in flat space one
can think of it as merely an exotic change of variables in the theory.
In this process a scalar anticommuting field is always produced
associated with a nilpotent supercharge $Q$. Furthermore, as argued in
\cite{conf,2dpaper} the twisted superalgebra implies that
the action rewritten in terms of these twisted fields is generically
$Q$-exact. In this case it is straightforward to construct a lattice
action which is $Q$-invariant provided {\it only} that we preserve
the nilpotency of $Q$ under discretization. 

In \cite{2dpaper} we gave a concrete application of these ideas by
constructing a gauge invariant lattice regularization of ${\cal N}=2$ super Yang-Mills
theory in two dimensions which was invariant under the twisted
scalar supersymmetry and an additional $SO(1,1)$ scaling symmetry.
In this paper we 
present results from the first numerical simulations of this model.

In this initial work we have examined a variety of
Ward identities following from the twisted supersymmetry. 
In the case of $U(2)$ we find
generally good agreement with theory for sufficiently large coupling
$\beta$.
Specifically we do not need to fine tune any additional couplings to
see a restoration of the twisted supersymmetry for small lattice spacing.
In the case of $SU(2)$ the agreement is even better -- we see no 
statistically significant breaking of the Ward identities even for
small $\beta$. Our results for the string tension indicate that the
lattice theory possesses a single confining phase. In addition our
initial
examination of the low lying spectrum appears to indicate the presence of
light scalar bound states.
 
We have also examined the issue of the phase of
the Pfaffian induced by integrating over the fermions. 
Our Monte Carlo simulations have been conducted in the phase
quenched ensemble in which this phase is neglected.
We present evidence that
the distribution of the phase
measured in the phase quenched ensemble peaks around zero 
at large coupling. This corresponds to our expectation that
the relevant Pfaffian is indeed real and
positive definite in the continuum
limit. Remarkably re-weighting
with the phase in the usual way does not appear to generate large
corrections to the phase quenched results at least in
the case of the supersymmetric Ward identities. We have also examined
the distribution of the eigenvalues of the scalar fields in the
theory.
The latter quantity yields
information of the nature of the quantum moduli space. We provide
strong 
evidence that the vacuum degeneracy of the classical 
theory is lifted at the quantum level. 

The paper
is organized as follows; we first summarize the reformulation of the
continuum theory is terms of twisted fields. This twisted theory has a natural
mapping to the lattice and we give the lattice action and the action of
the twisted supersymmetry on the lattice fields. The
details of our numerical algorithm are then described and lead to
a presentation
of our results both for $U(2)$ and $SU(2)$ gauge groups. 
A final section summarizes our conclusions and discusses
future work.  

\section{Continuum Twisted Formulation}
Consider a theory with ${\cal N}=2$ supersymmetry in two
dimensional Euclidean space. Such a theory contains
two Majorana supercharges $q^I_\alpha$ transforming under the global
symmetry group $SO(2)\times SO(2)_R$ where the subscript 
corresponds to two dimensional rotations and the superscript
describes the behavior under the R-symmetry corresponding to
rotating the two Majorana fields into one another.

The basic idea of twisting, which goes back to
Witten \cite{witten,twist_kawa,conf}, is to introduce a new rotation group
\[SO(2)^\prime={\rm diagonal\; subgroup}(SO(2)\times SO(2)_R)\]
and to decompose all fields now as representations of this new
rotation group. In simple terms what this means is that
whenever I do a rotation in the base space by some angle
I must do an equal rotation in the R-symmetry space. It
is equivalent to treating the two indices $I$ and $\alpha$ as
equivalent. Thus the supercharges are to be interpreted as
{\it matrices} in this twisted picture.
\beq q^{I}_\alpha\to q_{\alpha\beta}\eeq
Such a matrix may then be expanded on a basis of products
of 2D gamma matrices
\beq q=QI+Q_\mu\gamma_\mu+Q_{12}\gamma_1\gamma_2\eeq
The fields $(Q,Q_\mu,Q_{12})$ are called the {\it twisted supercharges}.
The original SUSY algebra then implies a corresponding twisted algebra
which takes the form
\beq
\{q,q\}_{\alpha\beta}=4\gamma^\mu_{\alpha\beta}p^\mu\\
\eeq
In components this reads
\begin{eqnarray}
\{Q,Q\}&=&\{Q_{12},Q_{12}\}=\{Q,Q_{12}\}=\{Q_\mu,Q_\nu\}=0\nonumber\\
\{Q,Q_\mu\}&=&P_\mu\nonumber\\
\{Q_{12},Q_\mu\}&=&-\epsilon_{\mu\nu}P_\nu
\end{eqnarray}
Notice that the momentum is $Q$-exact and this fact suggests that generically
both the entire energy-momentum tensor and hence the action will also
be $Q$-exact. 
Furthermore, it should be clear
the fermions admit a similar decomposition
\beq \Psi=\frac{\eta}{2}I+\psi_\mu\gamma_\mu+\chi_{12}\gamma_1\gamma_2\eeq
and the twisted theory will not contain spinors but antisymmetric
tensor fields. It is possible to abstract these p-form
fields and consider them as components of a single geometrical object -- a
\KD field.
\beq \Psi=(\frac{\eta}{2},\psi_\mu,\chi_{12})\eeq
A general complex \KD field $\Psi$ describes two Dirac spinors in
two dimensions. These spinors can be read off as the columns of
the original fermion matrix in a particular $\gamma$ matrix basis.
However if we think of the twisted component fields as real then 
$\Psi^\dagger=\Psi^T$ and the \KD field will describe two
Majorana fields as required by the ${\cal N}=2$ supersymmetry.   
Furthermore, any theory invariant under the scalar $Q$-symmetry
must contain commuting superpartners for these fermionic fields.
\beq \Phi=(\phib,A_\mu,B_{12})\eeq
The action for the Yang-Mills model may be written
as $S=\beta Q\Lambda$ where
\beq
\Lambda={\rm Tr}\int
d^2x\left(
\frac{1}{4}\eta[\phi,\phib]+2\chi_{12}F_{12}+
\chi_{12}B_{12}+\psi_\mu D_\mu \phib\right)\eeq
where each field is in the adjoint of some gauge group 
$X=\sum_{a=0}^{N_g-1} X^a T^a$  and we employ antihermitian generators $T^a$.
In practice we will consider both $U(N)$ and $SU(N)$ gauge groups with
the convention that in the former case
the traceless generators correspond to $a=0\ldots N_g-2$ and 
the final generator is proportional to the unit matrix $T^{N_g-1}\sim iI$.
The action of $Q$ on the fields is given by
\begin{eqnarray}
QA_\mu&=&\psi_\mu\nonumber\\
Q\psi_\mu&=&-D_\mu\phi\nonumber\\
Q\phi&=&0\nonumber\\
Q\chi_{12}&=&B_{12}\nonumber\\
QB_{12}&=&[\phi,\chi_{12}]\nonumber\\
Q\phib&=&\eta\nonumber\\
Q\eta&=&[\phi,\phib]
\end{eqnarray}
Notice that $Q^2$ generates an (infinitessimal) gauge transformation
parametrized by the field $\phi$. Carrying out the $Q$-variation
and subsequently integrating out the multiplier field $B_{12}$ we find
the on-shell action
\begin{eqnarray}
S&=&\beta {\rm Tr}\int d^2x\left(
\frac{1}{4}[\phi,\phib]^2-\frac{1}{4}\eta [\phi,\eta]-F_{12}^2\right.\nonumber\\
&-&D_\mu \phi D_\mu \phib-\chi_{12}[\phi,\chi_{12}]\nonumber\\
&-&2\chi_{12}\left(D_1\psi_2-D_2\psi_1\right)-2\psi_\mu D_\mu\eta/2\nonumber\\
&+&\left.\psi_\mu [\phib,\psi_\mu]\right)
\end{eqnarray}
The terms involving the twisted fermion fields
correspond to the component form of the \KD action
\beq \int \Psi .(d_A-d^\dagger_A)\Psi\eeq
where $d_A$ is the gauged exterior derivative and $\Psi$ is 
the real \KD field
introduced earlier (we will only consider flat space in this paper).

\section{Lattice Theory}
This continuum twisted action can be discretized according to the
prescription detailed in \cite{2dpaper}. The on-shell lattice
action is 
\begin{eqnarray}
S_L&=&\frac{\beta}{2}{\rm Tr}\sum_x \left(
\frac{1}{4}[\phi,\phib]^2+F^\dagger_{12} F_{12}\right.\nonumber\\
&-&\left.\frac{1}{4}\eta^\dagger[\phi,\eta]-\chi^\dagger_{12}[\phi,\chi_{12}]^{(12)}+
\psi^\dagger_\mu[\phib,\psi_\mu]^{(\mu)}\right.\nonumber\\
&+&\left.(D^+_\mu \phi)^\dagger D^+_\mu \phib-
2\chi^\dagger_{12}\left(D^+_1\psi_2 -D^+_2\psi_1\right)\right.\nonumber\\
&-&\left.2\psi^\dagger_\mu D^+_\mu\frac{\eta}{2}+{\rm h.c}\right)
\label{act}
\end{eqnarray}
where scalars such as $\phi(x)$ are associated with lattice sites,
vectors such as $U_\mu(x)$ with links and rank 2 tensors with
plaquettes. For example, $\chi_{12}(x)$ is a lattice field
associated with the $12$-plaquette at site $x$.
They are assigned corresponding gauge transformation properties
\cite{aratyn}:
\begin{eqnarray}
\phi(x)&\to& 
G(x)\phi(x)G^{-1}(x)\nonumber\\
U_\mu(x)&\to& G(x)
U_\mu(x)G^{-1}(x+\mu)\nonumber\\
\chi_{12}(x)&\to&G(x)\chi_{12}(x)G^{-1}(x+1+2)
\end{eqnarray}
where $G=e^{\phi}$ is a lattice gauge transformation.
Notice that for fields of non-zero spin the infinitessimal form of
this lattice gauge transformation naturally leads not to the
usual naive
commutator characteristic of adjoint fields
but to a {\it shifted} commutator. For link
fields this looks like
\beq [\phi(x)f_\mu(x)-f_\mu(x)\phi(x+\mu)]\eeq
In order to allow us to construct gauge invariant quantities
using the above gauge transformation rules it is
{\it essential} that each continuum field with {\it non-zero spin}
gives rise
to two lattice fields. 
This doubling of degrees of
freedom can be associated to the two possible orientations of
the underlying $p$-cube (for $p>0$) on which the field lives. 
The doubling can be
conveniently encompassed by promoting each field from real to
complex and assigning the conjugate fields to transform in the
following way
\begin{eqnarray}
U^\dagger_\mu(x)&\to& G(x+\mu)
U^\dagger_\mu(x)G^{-1}(x)\nonumber\\
\chi^\dagger_{12}(x)&\to&G(x+1+2)\chi^\dagger_{12}(x)G^{-1}(x)
\end{eqnarray}
This complexification has immediate consequences -- the gauge group
of the lattice theory is promoted from 
$U(N)$ to $GL(N,C)$ (or $SU(N)$ to $SL(N,C)$) and the usual
gauge links are no longer unitary matrices.
In addition to the fields we need to
define the lattice derivatives appearing in the
lattice action eqn.~\ref{act}.   
The action of the
gauge covariant forward difference operator acting on scalars
and vectors is defined by \cite{aratyn}
\begin{eqnarray}
D^+_\mu f(x)&=&U_\mu(x)f(x+\mu)-f(x)U_\mu(x)\nonumber\\
D^+_\mu f_\nu(x)&=&U_\mu(x)f_\nu(x+\mu)-f_\nu(x)U_\mu(x+\nu)
\end{eqnarray}
It clearly reduces to the usual
gauge covariant derivative acting on adjoint fields
in the naive continuum limit. Note that this derivative acting on a scalar
or site 
field yields a field which gauge transforms as a link field and
the corresponding derivative of a link field yields a field which
transforms like a tensor or plaquette field. 
These properties allow us to construct
a lattice analog of the continuum gauged exterior derivative
by appropriately anti-symmetrizing in the spacetime indices. It is
then possible to make a straightforward transcription of 
the continuum twisted
action to the lattice. 
Furthermore, it is possible to write a covariant backward
difference which is adjoint to the above operator for
gauge invariant quantities. Its action on link and plaquette fields is
given by
\begin{eqnarray}
D^-_\mu
f_{\mu}(x)&=&f_\mu(x)U^\dagger_\mu(x)-U^\dagger_\mu(x-\mu)f_\mu(x-\mu)\nonumber\\
D^-_\mu f_{\mu\nu}(x)&=&f_{\mu\nu}(x)U^\dagger_\mu(x+\nu)-U^\dagger_\mu(x-\mu)f_{\mu\nu}(x-\mu)
\end{eqnarray}
The action in eqn.~\ref{act} is invariant under the following lattice
supersymmetry transformation which
is a simple generalization of the continuum one given earlier
\begin{eqnarray}
QU_\mu&=&\psi_\mu\nonumber\\
Q\psi_\mu&=&-D^+_\mu\phi\nonumber\\
Q\phi&=&0\nonumber\\
Q\chi_{12}&=&B_{12}\nonumber\\
QB_{12}&=&[\phi,\chi_{12}]^{(12)}\nonumber\\
Q\phib&=&\eta\nonumber\\
Q\eta&=&[\phi,\phib]
\end{eqnarray}
where the continuum $D_\mu\phi$ is replaced by the lattice {\it forward}
difference $D^+_\mu\phi$ as required by
gauge invariance and the prime on the commutators reflects its shifted
nature as discussed earlier. 
The transformations of the conjugate fields are gotten by
taking the adjoint of these variations with the constraint that
$f^\dagger=-f$ for any scalar or site field. Notice that it is only for
the group $GL(N,C)$ that the adjoint of a link field 
can be taken as transforming
independently from the link field itself.
Finally the Yang-Mills
field strength appearing above is defined by
\beq
F_{\mu\nu}(x)=D^+_\mu U_\nu(x)\to F^{\rm cont}_{\mu\nu}\;{\rm as}\;a\to 0\eeq
Writing this term out we find
\beq
\beta {\rm Tr}\sum_x F^\dagger_{12}(x)F_{12}(x)\eeq
\beq
\beta{\rm Tr}\sum_x \left(2I-U_P-U^\dagger_P\right)+
\beta{\rm Tr}\sum_x \left(M_{12}+M_{21}-2I\right)
\eeq
where
\beq
U_P=U_1(x)U_2(x+1)U^\dagger_1(x+2)U^\dagger_2(x)\eeq
resembles the usual Wilson plaquette term
and 
\beq M_{12}=U_1(x)U^\dagger_1(x)U_2^\dagger(x+1)U_2(x+1)\eeq
is a new zero area Wilson loop term which would vanish if the
link variables were restricted to unitary matrices.
Notice the appearance of
the Wilson term depends crucially on the appearance of both
$F_{\mu\nu}$ and $F^\dagger_{\mu\nu}$ which lends some support to
the use of complex variables in the formulation of the theory.

The lattice action we have written down possesses one additional
$SO(1,1)$ symmetry corresponding to the transformations
\beq \psi_\mu\to \lambda\psi_\mu,\qquad\eta ,\chi_{12}\to
\frac{1}{\lambda}\eta, \chi_{12}\eeq
\beq
\phib\to \frac{1}{\lambda^2}\phib,\qquad\phi\to \lambda^2\phi\eeq
The transformation of the conjugate fields under this global
symmetry is identical. 
This symmetry is useful as it guarantees the absence of additive
mass renormalizations in the lattice theory.

Finally we should point out that the spectrum of this lattice
theory contains {\it no} lattice doubles
either fermionic or bosonic. This result follows from
the work of Rabin, Becher and Joos \cite{rabin,becher,joos} who show that
actions written in terms of exterior derivatives and tensor fields
may be discretized without generating doubled modes. 
The discretization prescription is given
explicitly by replacing the usual partial derivatives in the
continuum theory by appropriate difference operators:
\begin{center}
$D_\mu\to D^+$ if acts like $d$\\
$D_\mu\to D^-$ if acts like $d^\dagger$ 
\end{center}
This double free property can be seen explicitly in our case
by examining the form
of the fermion action.
Using the following decomposition of the \KD field
\beq
\Psi=\left(\begin{array}{c}
\eta/2\\
\chi_{12}\\
\psi_1\\
\psi_2\end{array}\right) 
\eeq
Our twisted fermion action can be recast in the form 
$\Psi^\dagger M(U,\phi)\Psi$ where the 
fermion operator $M$ is given in block form by
\beq
M=\left(\begin{array}{cc}
-[\phi,]^{(p)}&K\\
-K^\dagger&[\phib,]^{(p)}
\end{array}\right)\eeq
with the Yukawas lying along the diagonal and $K$ the 
lattice \KD operator taking the form
\beq
K=\left(\begin{array}{cc}
D^+_2&-D^+_1\\
-D^-_1&-D^-_2
\end{array}\right)\eeq
In the continuum theory the \KD field satisfies a reality condition 
$\Psi^\dagger=\Psi^T$ and integration over these anticommuting fields
yields the {\it Pfaffian} of the fermion operator ${\rm Pf}(M)$.
In the free limit we find that this prescription, when
applied to the above lattice operator, yields the determinant of
an explicitly double free lattice laplacian
${\rm Pf}(M)={\rm det}(K)={\rm det}(D^+_\mu D^-_\mu)$

As an aside we note that there is well known equivalence between
\KD fermions and staggered fermions -- the 4 component fields of
a single \KD field in two dimensions
can be mapped to site fields on a lattice of half the 
lattice spacing and the free \KD action goes over into the usual
staggered action. This is another way of understanding why
discretizations of the \KD theory avoid spectrum doubling. The
usual flavor replication of staggered fermions here becomes a bonus --
it yields an automatic description of the two degenerate fermions
required by ${\cal N}=2$ supersymmetry.
Of course it must be remembered that
the gauging of our lattice \KD action is not at all the
usual gauging of staggered fermions so the exact equivalence does
not persist in the interacting theory.

To conclude this description of the lattice theory we should return
to the issue of complexification.
The lattice formulation we have given requires a doubling of
degrees of freedom -- we have argued that this is quite natural
in any lattice theory and can be associated with the two possible
orientations of the underlying p-cube. We have parametrized this
doubling in terms of complex fields. However, the
target continuum theory that we are hoping to reproduce 
in the limit of vanishing lattice spacing corresponds to
putting the imaginary parts of the fields to zero or more
accurately to setting
\begin{eqnarray}
{\rm Im}X_\mu^a&=&0\;\; \mbox{all fields $X$ bar scalars}\nonumber\\ 
\phib&=&-\phi^\dagger
\end{eqnarray}
Let us examine what this means for the
complexified lattice theory. First, consider the effective
action that results from integrating out the grassmann variables.
While the complexified theory would lead to a determinant, integration
in the truncated theory should result in a Pfaffian. 
If we ignore possible phase
problems and replace the latter by a square root of
the determinant we can see that the fermion effective action
will still be gauge invariant after truncation to the real line.
The bosonic action also remains gauge invariant after
the projection and clearly both contributions target the correct
continuum theory in the classical continuum limit.
The remaining important issue is whether
the Ward identities corresponding to the twisted
supersymmetry still hold in the truncated theory
(or more conservatively, hold in the
limit of vanishing lattice spacing with no additional fine
tuning). We conjecture that this may be so and have followed this
approach so far in our numerical work.

It is possible to make some progress in understanding why this might indeed be
true. First parametrize the general $GL(N,C)$ link field $U_\mu(x)$ in terms
of a unitary component $u_\mu(x)$ and a positive definite hermitian
piece $R_\mu(x)$ in the following way
\beq U_\mu(x)=R_\mu(x)u_\mu(x)\eeq
Now consider the second term in the gauge action
\beq \beta{\rm Tr}\sum_x \left(M_{12}+M_{21}-2I\right)\eeq
and insert the general decomposition of the gauge link given above. The result
for $M_{12}$ is
\beq M_{12}=R^2_1(x)R^2_2(x+1)\eeq
with a similar result for $M_{21}$. Notice it is independent
of the unitary piece $u_\mu(x)$. 
Consider the theory in the continuum limit $\beta\to\infty$.
It should be clear that in such a limit
each $R_\mu(x)$ is driven to the identity and the complex bosonic
action smoothly approaches the usual one involving real fields. Furthermore,
for large $\beta$ the fermion operator is both independent of $R_\mu(x)$ and
antisymmetric. Thus in this limit the real and imaginary components of
the fermions decouple and the fermion determinant factors into the square
of a Pfaffian. This decoupling ensures that expectation values
of operators depending only on the real part of the fermion field
(the Majorana condition)
and computed for large $\beta$ 
will approach their values computed in the truncated theory.
Thus, the Ward
identities of the truncated theory should hold at least for large
$\beta$ since they can be viewed as coming from the complexified theory
(possessing explicit exact $Q$-symmetry)
in the limit of infinite $\beta$. 
Of course large $\beta$ also corresponds to the limit of vanishing lattice
spacing and we see that this line of reasoning constitutes an
argument that the 
Ward identities in the truncated lattice theory should be realized
without fine tuning in the continuum limit. As we shall show in the next
section our numerical results are consistent with this. Indeed, in the
case of $SU(2)$ the Ward identities appear to hold 
with small errors even for small $\beta$. 

\section{Simulations}
For our simulations we have taken the gauge links to be
unitary matrices taking values in either
$U(2)$ or $SU(2)$. As in the continuum
theory the scalars $\phi$ and $\phib$ are taken to be
complex conjugates of each other\footnote{the antihermitian nature of our basis
$T^a$ actually ensures that $\phi^\dagger=-\phib$}. 
In the case of $U(2)$ the bosonic action possesses an exact zero
mode $\phi_0=(0,0,0,\phi^3)$ which we hence regulate with the addition of
a mass term $m^2\sum_x \phib(x)\phi(x)$. At the end of the calculation we should
like to send $m\to 0$ to recover the correct target theory. 
The bosonic action is real positive semi-definite, 
gauge invariant and clearly has the
correct naive continuum limit. 

To this gauge and scalar action
we should add the effective action gotten by integrating
over the grassmann valued fields. We will represent this as
\beq {\rm det}^{\frac{1}{4}} (M(U,\phi)^\dagger M(U,\phi)+m^2)\eeq
where $M$ is the lattice \KD operator introduced earlier. The power
of $\frac{1}{4}$ reflects the Majorana nature of the 
continuum \KD field.
Notice that we have added a gluino mass term for the fermions
which regulates the corresponding fermion zero mode $\Psi_0=(0,0,0,\eta^3)$ 
arising in the $U(2)$ theory as a consequence of
supersymmetry. The $SO(1,1)$ symmetry in the 
massless case prohibits additive renormalization of this mass
as a result of quantum effects. 
In the case of $SU(2)$ this mass parameter can be set to zero.
Clearly, the form of the fermion effective action we employ does not
take into account any nontrivial phase associated with the
fermion determinant or Pfaffian - our simulations generate the
phase quenched ensemble.
We later examine the phase explicitly. 
  
To simulate this model we have used the RHMC algorithm
developed by Clark and Kennedy \cite{rhmc}.
The first step of this algorithm replaces the effective action
by an integration over auxiliary {\it commuting}
pseudofermion fields $F$, $F^\dagger$
in the following way
\beq {\rm det}^{\frac{1}{4}}(M^\dagger M+m^2)=\int DFDF^\dagger e^{-F^\dagger
\left(M^\dagger M+m^2\right)^{-\frac{1}{4}}F}\eeq
The key idea of RHMC is to use an optimal (in the minimax sense)
rational approximation to this inverse fractional power.
\beq\frac{1}{x^{\frac{1}{4}}}\sim \frac{P(x)}{Q(x)}\eeq
where 
\beq P(x)=\sum_{i=0}^{N-1}p_ix^i\qquad Q(x)=\sum_{i=0}^{N-1}q_ix^i\eeq
Notice that we restrict ourselves to equal order polynomials in numerator
and denominator.
In practice it is important to use a partial fraction representation
of this rational approximation
\beq\frac{1}{x^{\frac{1}{4}}}\sim
\alpha_0+\sum_{i=1}^{N}\frac{\alpha_i}{x+\beta_i}\eeq
The coefficients $\alpha_i$, $\beta_i$ for $i=1\ldots N$
can be computed offline using
the remez algorithm\footnote{many thanks to Mike Clark for providing us
with a copy of his remez code}. Furthermore, 
the coefficients
can be shown to be real positive. Thus the linear systems are well behaved
and unlike
the case of polynomial approximation the rational fraction approximations
are robust, stable and converge rapidly with $N$. 
The resulting pseudofermion action becomes
\beq S_{\rm PF}=\alpha_0 F^\dagger F+\sum_{i=1}^NF^\dagger
\frac{\alpha_i}{M^\dagger M+m^2+\beta_i}F\eeq
It is thus just a sum of standard 2 flavor pseudofermion actions
with varying amplitudes and mass parameters. In principle
this pseudofermion action can now be used in a conventional
HMC algorithm to yield an {\it exact} simulation of the original
effective action \cite{hmc}. 
This algorithm requires that we compute the pseudofermion
forces. For example, the additional force on the gauge links due
to the pseudofermions takes the form
\beq f_{U}=\frac{\partial S_{\rm PF}}{\partial U}=
-\sum_{i=1}^N\alpha_i\chi^{\dagger i}\frac{\partial}
{\partial U}\left(M^\dagger M\right)\chi^i\eeq
where the vector $\chi^i$ is the solution of the linear problem
\beq (M^\dagger M+m^2+\beta_i)\chi^i=F
\label{lin}\eeq 
The final
trick needed to render this approach feasible is to utilize a {\it multi-mass
solver} to solve all $N$ sparse linear systems simultaneously and with
a computational cost determined primarily by the smallest shift 
$\beta_i$. We use a multi-mass version of the
usual conjugate gradient CG algorithm \cite{multi}. In practice for the
simulations shown here we have used $N=15$ and an approximation
that gives an absolute bound on the relative error of $10^{-6}$ for
eigenvalues of $M^\dagger M$ ranging from $10^{-8}$ to $10$ 
which conservatively covers the range need for both our
$U(2)$ and $SU(2)$ runs. We monitor the spectrum continuously
to make sure our approximation remains good. Typically we have amassed
between $10^3$ and $10^4$ HMC trajectories for each set of
parameter values which leads to statistical errors of between
$0.1$ and $2.0$ percent depending on observable.

Finally we make some remarks on the representation of this fermion
operator.
For the purposes of computation this abstract lattice fermion
operator $M$ is replaced with a sparse matrix whose non-zero elements
are gotten by choosing an explicit basis for the group
generators and evaluating all traces over internal indices. For example
the term $\sum_x{\rm Tr}\psi_\mu^\dagger(x) D^+_\mu\eta(x)$ yields
\beq \psib_\mu^a(x) V_\mu^{ab}(x)\eta^b(x+\mu)-
\psib_\mu^a(x) V_\mu^{ba}(x)\eta^b(x)\eeq
where $V_\mu^{ab}(x)={\rm Tr}(T^a U_\mu(x)T^b)$. Similarly Yukawa
terms reduce to matrix elements of the form
\beq \etab^a(x)(\Phi^{ab}(x)-\Phi^{ba}(x))\eta^b(x)\eeq
where $\Phi^{ab}(x)={\rm Tr}(T^a \left(\sum_c\phi(x)^cT^c\right) T^b)$.
In practice we use the basis
\beq T^0=i\sigma_1\qquad T^1=i\sigma_2\qquad T^2=i\sigma_3\eeq
where $\sigma_i$ are the usual Pauli matrices
and in the case of $U(2)$ the generator $T^3=iI$

As usual the running time of the simulations is dominated by 
the need to solve the
linear system eqn.~\ref{lin} for each step down a Monte Carlo
trajectory. We use
the usual sparse matrix techniques to optimize the CG-solver to
accomplish this.  
\section{Phase quenched U(2) model}
The numerical results we present here come from
simulations where the lattice length takes the values
$L=2,4,8$ for mass parameter $m=0.1$ and $L=3,6$ for $m=0.01$.
The coupling varies over the range $\beta=0.5\to 4.0$. 
Clearly Ward identities corresponding to the twisted supersymmetry
$Q$ are of prime interest. They are simply expectation
values of the form $<QO>$ and should be zero by supersymmetry.
Perhaps the simplest of these corresponds to the action itself
$<S=Q\Lambda>=<S_{\rm B}>+<S_{F}>=0$.
This fact, together with the quadratic nature
of the fermion action 
allows us to compute the bosonic (gauge plus scalar) 
action exactly using a simple scaling argument and for {\it all}
values of the coupling constant $\beta$ we find
\beq \beta<S_B>=\frac{3}{2}N_gL^2\eeq
where $N_g=4$ is the number of generators of $U(2)$. 
Figure 1. shows a plot of $\frac{\beta}{6L^2}<S_B>$ for a range of
couplings $\beta$ and three different lattice sizes using a mass $m=0.1$.
The bold lines shows the analytic prediction based on supersymmetry (for
clarity we have added multiples of $-0.25$ to the curves and lines to
split up the data from different lattice sizes)
While there are clearly deviations of order 4-5\% at small coupling these
appear to disappear at large $\beta$ in line with our
expectations. 
Figure 2. shows equivalent data for $L=3,6$ at
the smaller gluino mass $m=0.01$. Again the horizontal lines show
the analytic result expected from supersymmetry. In this case the
deviations for the small lattice appear larger at small coupling but
nevertheless appear to converge toward the theoretical expectation
on the basis of supersymmetry as $\beta$ is increased. The larger 
lattice $L=6$ data are even better.
We have also examined other Ward identities corresponding to the local
operator
choices 
\begin{eqnarray}
O&=&O_1(x)=\etab(x)[\phi(x),\phib(x)]\nonumber\\
O&=&O_2(x)=\chib_{\mu\nu}F_{\mu\nu}\nonumber\\
O&=&O_3(x)=\psib_\mu(x) D^+_\mu\phib(x)
\end{eqnarray}
After $Q$-variation we find
\begin{eqnarray}
QO_1&=&[\phi,\phib]^2-\etab[\phi,\eta]\nonumber\\
QO_2&=&F^\dagger_{\mu\nu}(x)F_{\mu\nu}(x)-
\chib_{\mu\nu}D^+_{\left[\mu\right.}\psi_{\left.\nu\right]}\nonumber\\
QO_3&=&-D^+_\mu\phi D^+_\mu\phib-\psib_\mu D^+_\mu\eta-\psib_\mu[\psib_\mu,\phi]
\end{eqnarray}
The results for expectation values of these $Q$-variations
for $\beta=4.0$ and $m=0.01$ are displayed in tables 1. and 2. corresponding to
lattice sizes $L=3$ and $L=6$ respectively
(we denote the bosonic contribution to the Ward identity by $B$ and
the fermionic one by $F$).
\begin{table}
\begin{center}
\begin{tabular}{||l|l|l|l||}
\hline
$O$   & $B$          & $Re(F)$        & $Im(F)$          \\\hline
$O_1$ & $0.88(7)  $ & $-0.86(6) $   & $-0.0003(5)$   \\\hline
$O_2$ & $0.285(16)$ & $-0.297(16)$   & $-0.0009(28)$      \\\hline
$O_3$ & $0.90(45)$ & $-0.986(4)$   & $0.0035(39)$      \\\hline
\end{tabular}\end{center}
\caption{Ward identities for $U(2)$ and $\beta=4.0$,$L=3$}\end{table}
\begin{table}
\begin{center}
\begin{tabular}{||l|l|l|l||}
\hline
$O$   & $B$          & $Re(F)$        & $Im(F)$       \\\hline
$O_1$ & $0.38(8) $ & $-0.43(10)$  & $-0.0008(15)$  \\\hline
$O_2$ & $0.409(22) $ & $-0.386(25) $  & $-0.0006(7)$    \\\hline
$O_3$ & $1.22(49)$ & $-0.995(1) $  & $-0.0003(20)$  \\\hline
\end{tabular}\end{center}
\caption{Ward identities for $U(2)$ and $\beta=4.0$,$L=6$}\end{table}
Notice that the imaginary part of the fermionic correlator is always
small and statistically consistent with zero\footnote{the fermionic correlator
$<\Psi_i\Psi_j>=\frac{1}{2}M^{-1}_{ij}$ where spacetime, group and \KD indices
are combined into a single index $i$ and the factor of $\frac{1}{2}$ originates
from the Majorana nature of the \KD field}.
Within the statistical errors the bosonic and fermionic contributions
do add to zero confirming the presence of the $Q$-symmetry in the
quantum lattice theory for this coupling.

The apparent breaking of supersymmetry at small $\beta$ appears to be correlated to
the symmetry properties of the lattice fermion operator $M(U,\phi)$.
In the continuum this operator is complex antisymmetric and hence its
determinant can be written as the square of a Pfaffian. This
is true for both $SU(2)$ and $U(2)$ theories (the $U(1)$ or trace mode of the
scalars and gauge field disappear from the fermion operator for
all couplings rendering both continuum theories equivalent in this
respect).

The antisymmetric condition is very important also in the lattice theory
because in this
case we have argued that the real and imaginary components of the
fermion decouple - a necessary condition for existence of
a Pfaffian and the truncation of the theory
to the real line. If the matrix is not antisymmetric this factorization
cannot be achieved and supersymmetry must necessarily be broken.
In the case of $SU(2)$ it is not hard to see that the matrix is indeed
antisymmetric and we exploit this fact later when we examine the phase of
the Pfaffian for the $SU(2)$ system. This property is not shared by the
$U(2)$ model in general though since the trace degrees of freedom of the
gauge link do not decouple for finite $\beta$. Approximate decoupling
will now occur in the region of large coupling $\beta$ where
the approximation $U_\mu(x)=1+A_\mu(x)+\cdots$ becomes accurate.
By examining the fourth root of the plaquette we learn that $|A_\mu|\sim 0.05$
for $\beta=3.0$. Hence corrections to this approximation will be
less than one percent if $\beta>3$ consistent with the restoration of
supersymmetry we see in the large $\beta$ regime.
This decoupling can be seen explicitly in figure 3. which shows the
Monte Carlo evolution of the scalar field for $\beta=3.0$, $L=6$
and $m=0.01$ in the $U(2)$ theory. Two
modes are shown corresponding to $\phi^3$ (the trace mode) and $\phi^0$
a traceless mode.
The trace mode $\phi^3$ behaves as a quasi massless degree of
freedom undergoing large
fluctuations regulated only by the imposed IR cut-off of $m=0.01$ while
the traceless degrees of freedom fluctuate 
independently over a scale two orders of magnitude
smaller. 
The spectrum of a typical equilibrated configuration for $\beta=4.0$
on a lattice of size $L=6$
is shown in figure 4. Much of the spectrum is concentrated close to
the imaginary axis and is indicative of light continuum-like states.
Notice the approximate pairing of eigenvalues $(\lambda,-\lambda)$
related to the approximate antisymmetry of the fermion operator at
this large coupling.
There are in addition 
``islands'' of additional states at large eigenvalue which we
conjecture
are related to the large fluctuations of the nonzero
trace modes of the scalars. These don't appear to have any continuum
interpretation. 
\section{Phase quenched SU(2) model}
Since the $SU(2)$ theory contains no exact zero modes we have been
able to simulate the model at exactly zero gluino mass. Figure 5.
shows a plot of the mean bosonic action normalized to unity as a function
of $\beta=0.5\to 4.0$ for $L=3,6,8$ for the $SU(2)$ theory.
In contrast with the $U(2)$ lattice theory the scalar
supersymmetry appears
to be good here down to small coupling $\beta$. As we have remarked we
conjecture that this is related to the presence of an exact
antisymmetry of $M$ in the $SU(2)$ case.
Clearly the Yukawas possess this property. What is non trivial is that
this is also true of the gauged lattice \KD term. In general this
term is antihermitian but in the case of $SU(2)$ it is also
{\it real}. This follows from the special property of
the Pauli matrices 
\beq
e^{\mu_i\sigma_i}=\cos{\left|\mu \right|}I+
\frac{\sin{\left|\mu\right|}}{\left|\mu\right|}\mu_i\sigma_i\eeq
Using this representation it is easy to show that $V_{\mu}^{ab}(x)$ is real.
Figure 6. shows a plot of the eigenvalues of the $SU(2)$ theory
for $L=6$ at both $\beta=0.5$ and $\beta=4.0$ in which the
exact pairing of eigenvalues is manifest. Notice that as $\beta$ increases
the real parts of these eigenvalues decrease. In the limit we expect
the eigenvalues to lie along the imaginary axis yielding a real, positive
definite determinant as for the continuum theory.
We have additionally measured the same local Ward identities as for $U(2)$
with the results listed in tables 3. and 4. Here, the data is taken from
runs with $L=6$ and two different values of the coupling $\beta=0.5$
and $\beta=4.0$. 
\DOUBLETABLE{
\begin{tabular}{||l|l|l|l||}
\hline
$O$   & $B$          & $Re(F)$        & $Im(F)$          \\\hline
$O_1$ & $4.70(20)$ & $-4.72(04) $   & $-0.04(04)$   \\\hline
$O_2$ & $1.95(07)$ & $-2.02(02)  $   & $0.014(14) $      \\\hline
$O_3$ & $5.76(15) $ & $-5.74(05) $   & $-0.043(35)$      \\\hline
\end{tabular}}
{
\begin{tabular}{||l|l|l|l||}
\hline
$O$   & $B$          & $Re(F)$        & $Im(F)$       \\\hline
$O_1$ & $0.167(10)$ & $-0.195(8)$    & $-0.007(07)$  \\\hline
$O_2$ & $0.344(10) $ & $-0.346(1) $   & $-0.0007(8)$    \\\hline
$O_3$ & $0.759(21) $ & $-0.739(4) $   & $0.0012(17)$  \\\hline
\end{tabular}}
{Ward identities for $SU(2)$ and $\beta=0.5$,$L=6$}
{Ward identities for $SU(2)$ and $\beta=4.0$,$L=6$}
Again for both small and large coupling these local Ward
identities appear to be satisfied to within statistical error.

Since the simulations of the $SU(2)$ are carried out at zero mass we
have a priori no lower bound on the eigenvalue spectrum and so we
monitor the smallest eigenvalue continuously to ensure it lies within
the limits required by our minimax approximation to the inverse
fourth root of the fermion operator\footnote{For the $U(2)$ theory
this is rigorously bounded below by m}. A typical plot of the Monte Carlo
evolution of $|\lambda_{\rm min}|$ is shown in figure 7. We observe that
the magnitude of this smallest eigenvalue decreases with
increasing $\beta$ and $L$ but we see no evidence for an exact
zero mode as would be expected for a supersymmetric theory whose
flat directions survive quantum corrections. We will return to this
issue when we discuss the quantum moduli space.

Since we employ periodic boundary conditions the partition function
we simulate
yields the Witten index of the theory and is explicitly independent of
coupling constant (recall that $\frac{\partial lnZ}{\partial\beta}=<S>=0$).
This in turn implies that there can be no thermodynamic singularity for
finite $\beta$ and we expect the theory exists in a single phase.
Figure 8. confirms this expectation by plotting
the string tension as estimated from the $2\times 2$ Creutz ratio
as a function of $\beta$ for a lattice of size $L=6$. The string
tension appears to be non-zero and
smoothly varying over this range of coupling.

We have also examined a couple of correlation functions which give us
direct access to the low-lying mass states of the theory.
The simplest $SO(1,1)$ and gauge invariant bosonic correlator takes the
form 
\beq G_B(t)=\sum_{x,x^\prime}
<\phi^a(x,t)\phib^a(x,t)\phi^b(x^\prime,0)\phib^b(x^\prime,0)>\eeq
and the sum over
spatial sites $x,x^\prime$ projects to the zero momentum sector.
We have also examined a fermionic correlator of the form
\beq G_F(t)=\sum_{x,x^\prime}<\etab^a(x,t)\phi^a(x,t)
\eta^b(x^\prime,0)C^b(x^\prime,0)>\eeq
where $C=[\phi,\phib]$.
These functions are shown in figure 9. together with fits to hyperbolic
cosines.
As a consequence of supersymmetry we expect that the lowest lying bosonic
and fermionic states should have the same mass. Within statistical errors
this is true (and indeed this state is rather light). However the
errors on the fermion are large and the current data is really
inadequate to decide this question. Of course it is also not clear we
have the correct interpolating operator for the lightest fermion state -- 
further investigations of these
issues are underway.
\section{Reality of the fermionic effective action}
Up to this point we have neglected a possible phase associated with
the Pfaffian induced by integration over the fermion fields.
As we have seen the truncation to the real line in general breaks
the supersymmetry in the $U(2)$ model so we will concentrate on the
$SU(2)$ case where the fermion operator is an antisymmetric matrix and
a Pfaffian can be unambiguously defined. As usual we can always
compensate for neglecting this phase in the Monte Carlo simulation
by re-weighting all observables by
the phase factor according to the simple rule
\beq
<O>=\frac{<Oe^{i\alpha(U,\phi)}>_{\rm \alpha=0}}{<e^{i\alpha(U,\phi)}>_{\rm
\alpha=0}}\eeq
We thus have computed the Pfaffian as one of our observables allowing us
to carry out this re-weighting procedure when computing expectation values.
The Pfaffian computation is carried out by using a variant of
Gaussian elimination with full pivoting to transform the $2n\times 2n$
dimensional antisymmetric matrix $M$ into
the canonical form
\beq
\left(\begin{array}{ccccc}0 &\lambda_1&0 &0&\ldots \\
                         -\lambda_1 & 0 &0 &0&\ldots \\
			  0 & 0 & 0 &\lambda_2&\ldots\\
			  0 & 0 & -\lambda_2 & 0&\ldots\\
			  0 & 0 & 0 & 0 &\ldots\\
      \end{array}\right)\eeq
Then 
\beq {\rm Pf}(M)=\prod_{i=1}^n \lambda_i\eeq
Consider first the phase itself. Figure 10. shows a plot 
of the distribution of $\cos{\alpha}$ for $L=3$ and $\beta=4.0$.
A strong peak close to $\alpha=0$ is manifest. This peak strengthens
with increasing $\beta$ as the scalars (which are responsible for
a non-zero phase) are driven closer to zero. Figure 11. shows a plot
of $\cos{\alpha}$ vs $\beta$ for $L=6$
($<\sin{\alpha}>$ is small and always statistically
consistent with zero).
We see that it increases from values close to zero
to attain
\beq <\cos{\alpha}>\sim 0.2 \qquad \beta=4.0\eeq
Hence, in the
range of coupling we have simulated, it clearly
fluctuates strongly from the naive value of unity used in generating the phase
quenched ensemble. 
In light of this we have re-examined the Ward identities now weighted
with this phase factor. 
Table 5. shows the mean re-weighted bosonic action
together with the phase quenched value for $L=3$ and all $\beta$
(note that these numbers are not normalized to unity as in the earlier
plots).
While re-weighting typically amplifies the estimated error it does
not appear to change the mean value for this observable at least
within the statistical errors. 
\TABLE{
\begin{tabular}{||l|l|l||}
\hline
$\beta$   & $S_B$   & ${\rm Reweighted}\;S_B $          \\\hline
$0.5$ & $40.35(19)$ & $39.80(95)$    \\\hline
$1.0$ & $40.45(18)$ & $40.42(117)$     \\\hline
$2.0$ & $40.11(32)$ & $40.80(130)$        \\\hline
$2.5$ & $40.16(32)$ & $40.62(130)$\\\hline
$3.0$ & $39.74(25)$ & $40.11(140)$\\\hline
$3.5$ & $39.99(30)$ & $40.58(200)$\\\hline
$4.0$ & $39.73(30)$ & $39.66(220)$\\\hline
\end{tabular}
\caption{Ward identities for $SU(2)$ and $\beta=4.0$,$L=6$}}
This conclusion is strengthened by examining other re-weighted Ward
identities corresponding to the set of
operators $O_1$, $O_2$ and $O_3$ given earlier.
Tables 6. and 7. compare the naive (phase quenched)
expectation values with their re-weighted values for lattice size
$L=6$ and $\beta=0.5$ and $\beta=4.0$. Again, there is no evidence that
the central values change within the (admittedly large) statistical
errors. 
\begin{table}
\begin{center}
\begin{tabular}{||l|l|l|l|l||}
\hline
$O$   & $B$        & $Re(F)$      & Reweighted $B$ & Reweighted $Re(F)$ \\\hline
$O_1$ & $4.70(20)$ & $-4.72(03) $ & $6.46(210)$    &  $-5.04(100)$\\\hline
$O_2$ & $1.95(07)$ & $-2.02(02) $ & $2.14(100)$   &   $-2.01(50) $ \\\hline
$O_3$ & $5.76(15)$ & $-5.74(02) $ & $5.74(200) $   &  $-5.25(200) $ \\\hline
\end{tabular}\end{center}
\caption{Naive vs reweighted Ward identities for $SU(2)$ and $\beta=0.5$,$L=6$}\end{table}

\begin{table}
\begin{center}
\begin{tabular}{||l|l|l|l|l||}
\hline
$O$   & $B$        & $Re(F)$      & Reweighted $B$ & Reweighted $Re(F)$ \\\hline
$O_1$ & $0.167(10)$ & $-0.195(7)$ & $0.159(40)$    &  $-0.137(50)$\\\hline
$O_2$ & $0.344(10)$ & $-0.346(1)$ & $0.356(60)$   &   $-0.348(50)$ \\\hline
$O_3$ & $0.759(21)$ & $-0.739(4)$ & $0.717(110)$   &  $-0.733(120)$\\\hline
\end{tabular}\end{center}
\caption{Naive vs reweighted Ward identities for $SU(2)$ and $\beta=4.0$,$L=6$}\end{table}
Taken at face value 
this apparent weak dependence of the expectation values on reweighting
seems to indicate that the phase fluctuates approximately independently of
the other observables leading to an, at least approximate, factorization in
the reweighted observable
\beq <O>_{\rm full}=
\frac{<Oe^{i\alpha}>_{\alpha=0}}{<e^{i\alpha}>_{\alpha=0}}\sim
\frac{<O>_{\alpha=0}<e^{i\alpha}>_{\alpha=0}}
{<e^{i\alpha}>_{\alpha=0}}=<O>_{\alpha=0}
\eeq
As a practical matter this means that expectation values obtained
within the phase quenched approximation may be quite reliable in spite
of the large phase fluctuations. 
\section{Quantum moduli space}
Finally we turn to an important issue concerning the two scalar fields
that appear in this theory. The classical vacua allow for
any set of scalars which are constant over the lattice and
satisfy 
\beq[\phi,\phib]=0\eeq
In the case of $SU(2)$ and using the parameterization $\phi=\phi_1+i\phi_2$
we find vacuum states of the form 
\beq \phi_1=(A,0,0)\qquad \phi_2=(B,0,0)\eeq
together with
global $SU(2)$ rotations of this configuration.
Thus we have a classical vacuum state for any value of
$A$ and $B$. This space of vacuum solutions is referred to
as the moduli space of the theory. The presence of such a non-trivial
moduli space corresponds to the existence of flat directions in the theory.
This is problematic in the quantum theory as integration over such
flat directions may induce IR divergences. However we find that this
is not the case in practice -- the quantum ground state appears to
be unique and the flat directions are lifted
(except for the trivial $U(1)$ factor associated with trace part of
$U(2)$). This can be seen from
the distribution of eigenvalues of the scalars averaged over the lattice.
Figure 12. shows a plot of this distribution for the $SU(2)$ theory on a $L=3$
lattice both for $\beta=0.5$ and $\beta=4.0$.
The distribution is symmetric about the origin so we show only the
positive values here.
Both for small and large coupling the distribution possesses a well
defined peak with a tail extending out to large eigenvalue. 
Notice that the peak moves to smaller values as
$\beta$ increases in line with the observed suppression
of scalar field fluctuations with increasing coupling. The data
indicates that the
partition function at least exists and most likely at least some
of the lower moments of the scalar field. This result is
reminiscent of similar results obtained for zero dimensional
supersymmetric Yang-Mills integrals \cite{staud}. 

\section{Conclusions}
In this paper we have presented initial results from a full simulation
of the ${\cal N}=2$ super Yang-Mills theory in two dimensions.
The lattice action we employ was derived in \cite{2dpaper} and follows
from a reformulation of the theory in terms of {\it twisted} fields.
It is invariant under a global $SO(1,1)$ symmetry and $U(N)$
lattice gauge transformations and exhibits, at least in its
complex form, an exact scalar supersymmetry.
We show results for both $U(2)$ and $SU(2)$ theories for a range
of lattice size $L$ and coupling $\beta$. To check for
supersymmetry we have examined a number of Ward identities.

In the case of $U(2)$ the supersymmetry
appears to be only exact at large $\beta$. At small $\beta$
the fermion operator is not antisymmetric, its
Pfaffian is not even defined and the truncation of the
complexified theory to the real line appears to break supersymmetry.
However,
the theory does not seem to need any fine tuning to regain
supersymmetry for large coupling $\beta$ and hence in the continuum
limit.

In the case of $SU(2)$ the lattice gauged \KD operator is
real and antisymmetric and supersymmetry is manifest for
all couplings - we see no statistically significant violation of
any Ward identities at the 1\% level for any coupling or
lattice size. We conjecture that the key property which
allows the supersymmetry to be realized is the antisymmetric
property of the fermion operator. Since the latter is
always antihermitian this in turn boils down to a reality
property on the gauged \KD operator.
At first glance it appears that
$SU(2)$ is rather special in this regard. However, reality of the $SU(N)$
theories will be guaranteed for all $N$ if instead of using the
$N\times N$ basis for the generators we employ the
basis constructed from the structure constants themselves
$T^a_{bc}=f_{abc}$. Such a choice yields the same
naive continuum limit but guarantees that the gauged
\KD action is real and antisymmetric.
We conjecture that such models will resemble
$SU(2)$ and exhibit an exact scalar supersymmetry at the
quantum level.

In the case of $SU(2)$ we have also examined the validity of
the phase quenched approximation used in our simulations. The
phase appears to approach zero in the large $\beta$ continuum limit
as naively expected. Remarkably, re-weighting 
observables with the phase does not
appear to have a strong influence on expectation values even for small
$\beta$ at least
in the case of observables corresponding to Ward identities.
We also show results on the distribution of the eigenvalues of the
scalar fields. This distribution possess a well-defined
peak which narrows and moves to smaller values as
$\beta$ increases. This structure is similar to the case of zero dimensional
SUSY Yang-Mills and indicate that the classical vacua are lifted via
quantum effects. 

To summarize, our initial numerical investigations of twisted
formulations of lattice super
Yang-Mills theories are quite positive -- it appears that 
full dynamical simulations of these theories are possible 
with rather moderate computational resources (the work presented here
was obtained with $O(300)$ single CPU days on a P3 cluster). 
These preliminary results provide evidence that supersymmetry
is indeed realized at the quantum level and have already allowed us
to check certain qualitative features of the theory -- for example,
our data
support the lifting of the classical
vacua and finiteness of the partition function. They also support a single
phase picture for this theory.
It would be nice to extend these calculations to larger lattices
to be able to get more reliable results for the low-lying spectrum
which could be compared to the SDLCQ results reported in
\cite{n=2LC} and to the dimensionally reduced quantum mechanics
case \cite{wos}.
It is important also to examine Ward identities corresponding to
other elements of the twisted supersymmetry to see whether indeed
the other supersymmetries are realized without fine tuning in the
continuum limit. Results from these investigations will be published
elsewhere \cite{newish}.

It should be stressed that low dimensional super Yang-Mills theories are
of great interest because of their conjectured connections to
various types of (super)gravity theory. They are a place where
lattice simulations could potentially play an important role since
high precision
exact dynamical simulations are possible on large lattices.
In principle, the strongest connections to gravitational systems
are exhibited for theories with sixteen supercharges rather than
the four supercharge case considered here. However, dimensional
reduction of the ${\cal N}=D=4$ lattice action constructed in
\cite{4dpaper} would yield $Q$-exact actions for these systems
whose fermion content could be represented using \KD fields. 
Notice that although the \KD action is related to the usual
staggered fermion action there is no ``fourth root'' problem with
these theories since the fermion degeneracy associated with these
lattice actions precisely accounts for the number of
physical fermions required 
by the extended supersymmetry.

\acknowledgments
This work
was supported in part by DOE grant DE-FG02-85ER40237. The author would
like to thank Mithat Unsal and Toby Wiseman for useful discussions.

\vfill
\newpage

\EPSFIGURE{actu2m1.eps, width=5.0in}
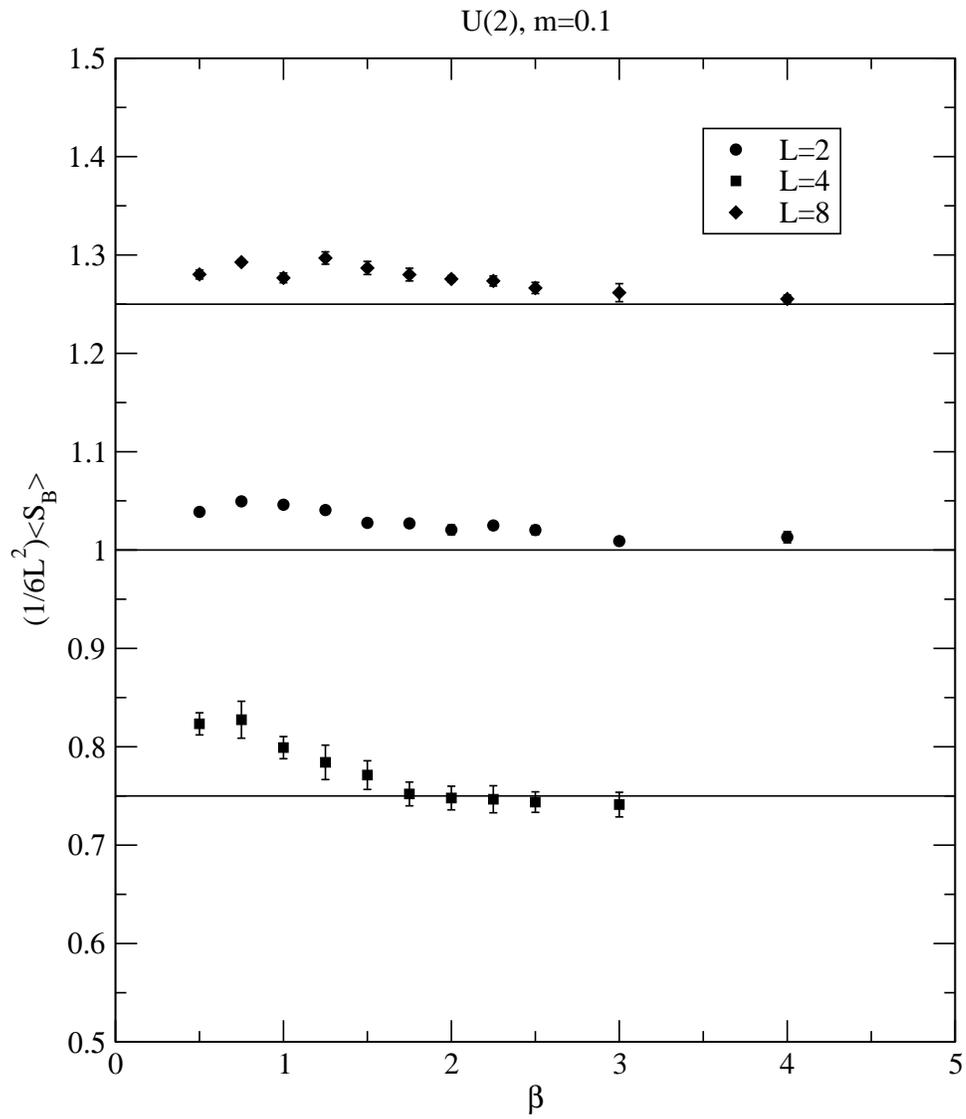
{$\frac{\beta}{6L^2}<S_B>$ for $U(2)$ at $m=0.1$}

\EPSFIGURE{actu2m01.eps, width=5.0in}
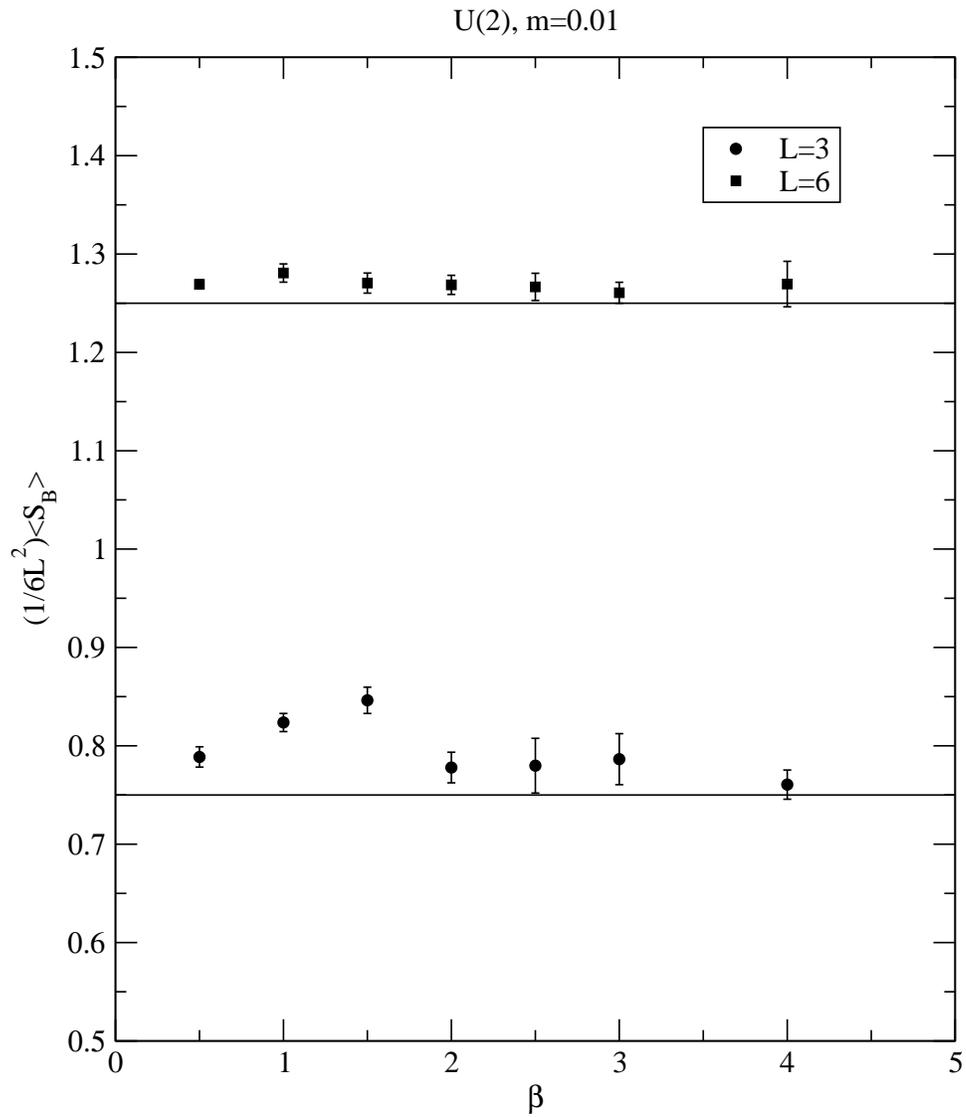
{$\frac{\beta}{6L^2}<S_B>$ for $U(2)$ at $m=0.01$}

\EPSFIGURE{scalar_u_m01_l6.eps, width=5.0in}
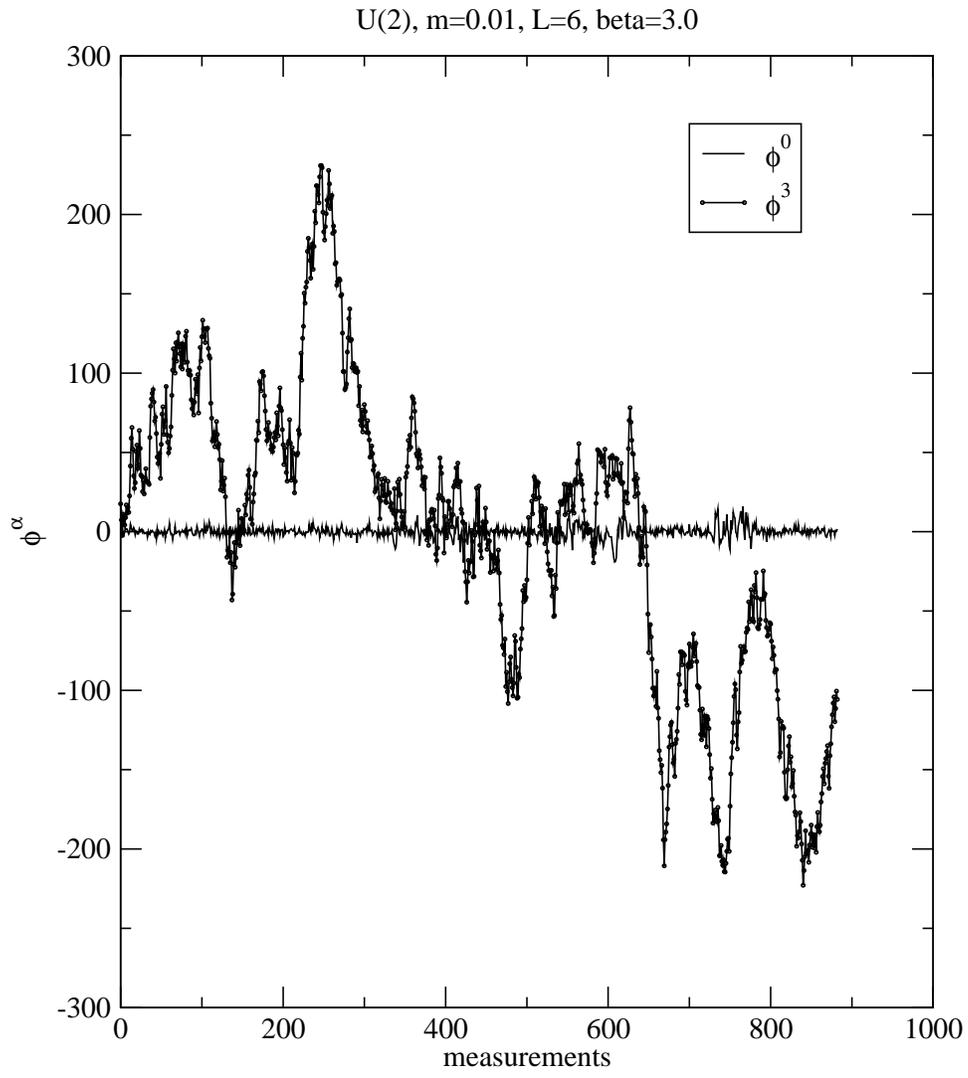
{Monte Carlo evolution of trace and traceless components of scalars}

\EPSFIGURE{eigenu2b4.eps, width=5.0in}
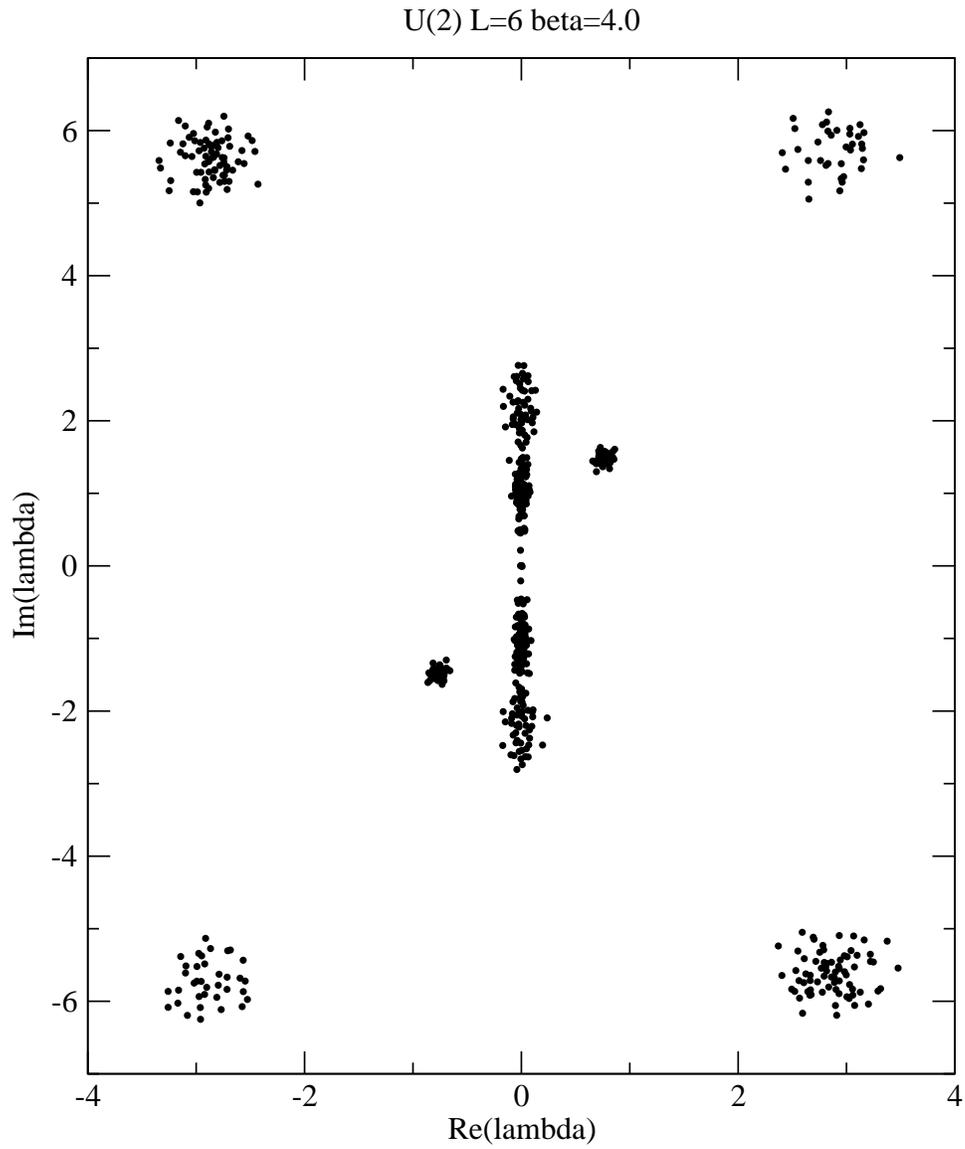
{Eigenvalue spectrum for $U(2)$ theory at $\beta=4.0$}

\EPSFIGURE{actsu2.eps, width=5.0in}
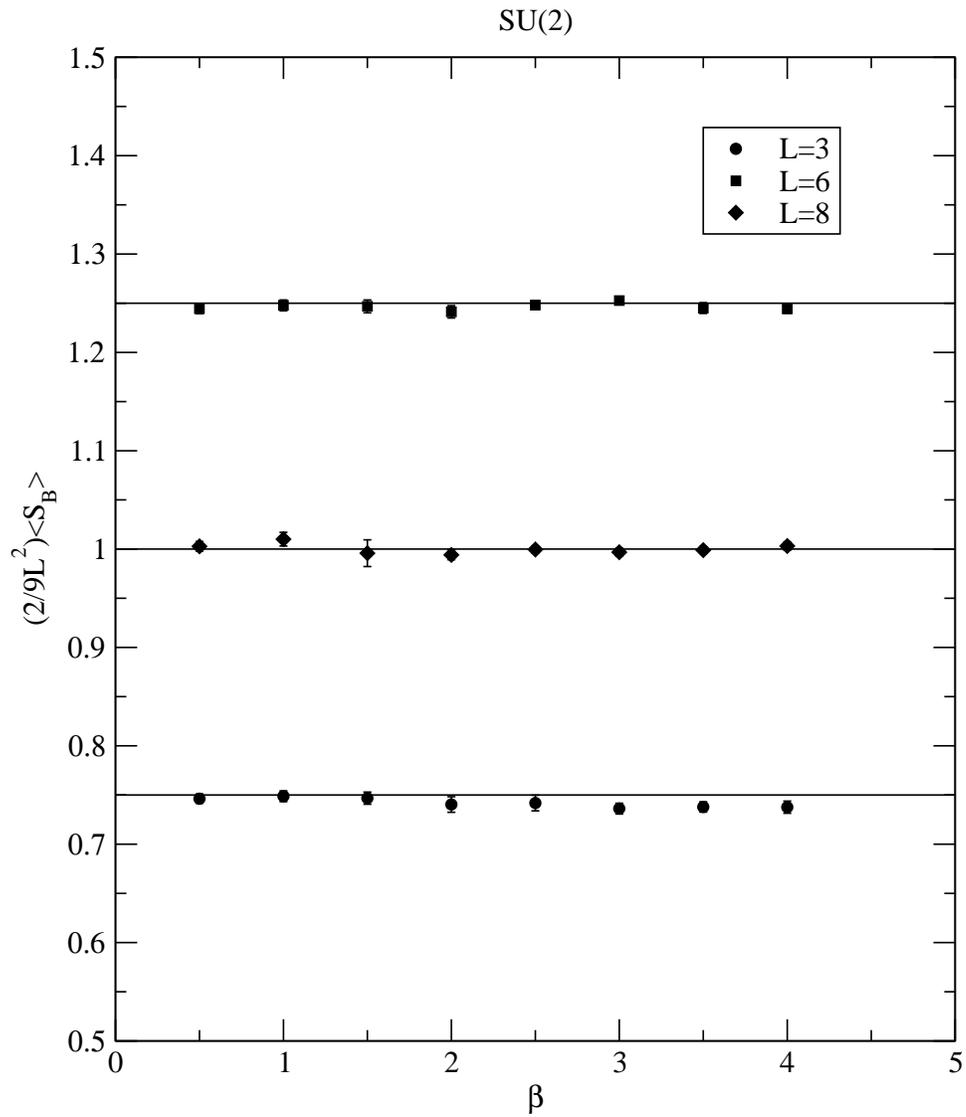
{$\frac{2\beta}{9L^2}<S_B>$ for $SU(2)$}

\EPSFIGURE{eigensu2all.eps, width=5.0in}
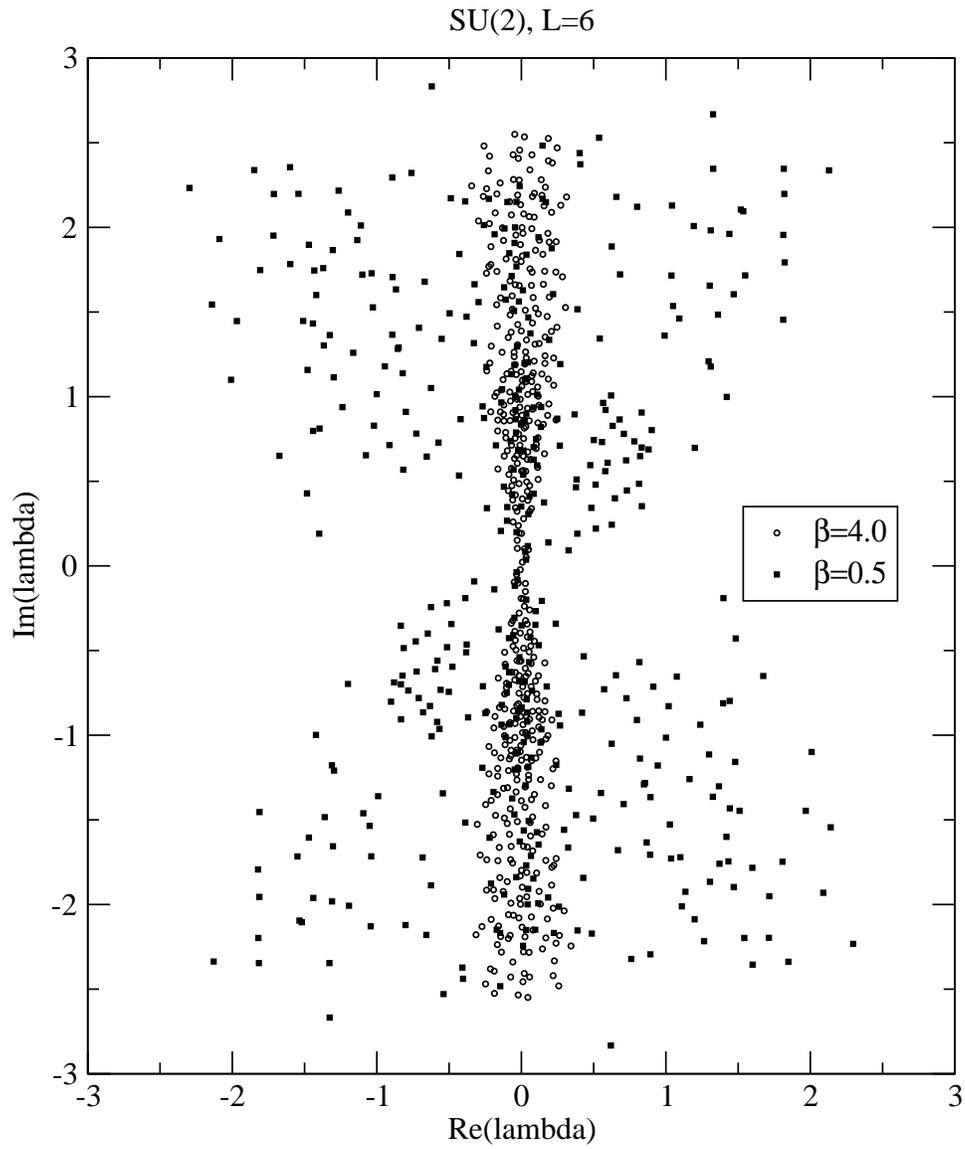
{Eigenvalue spectrum for $SU(2)$ theory at $L=6$}

\EPSFIGURE{lambda_small.eps, width=5.0in}
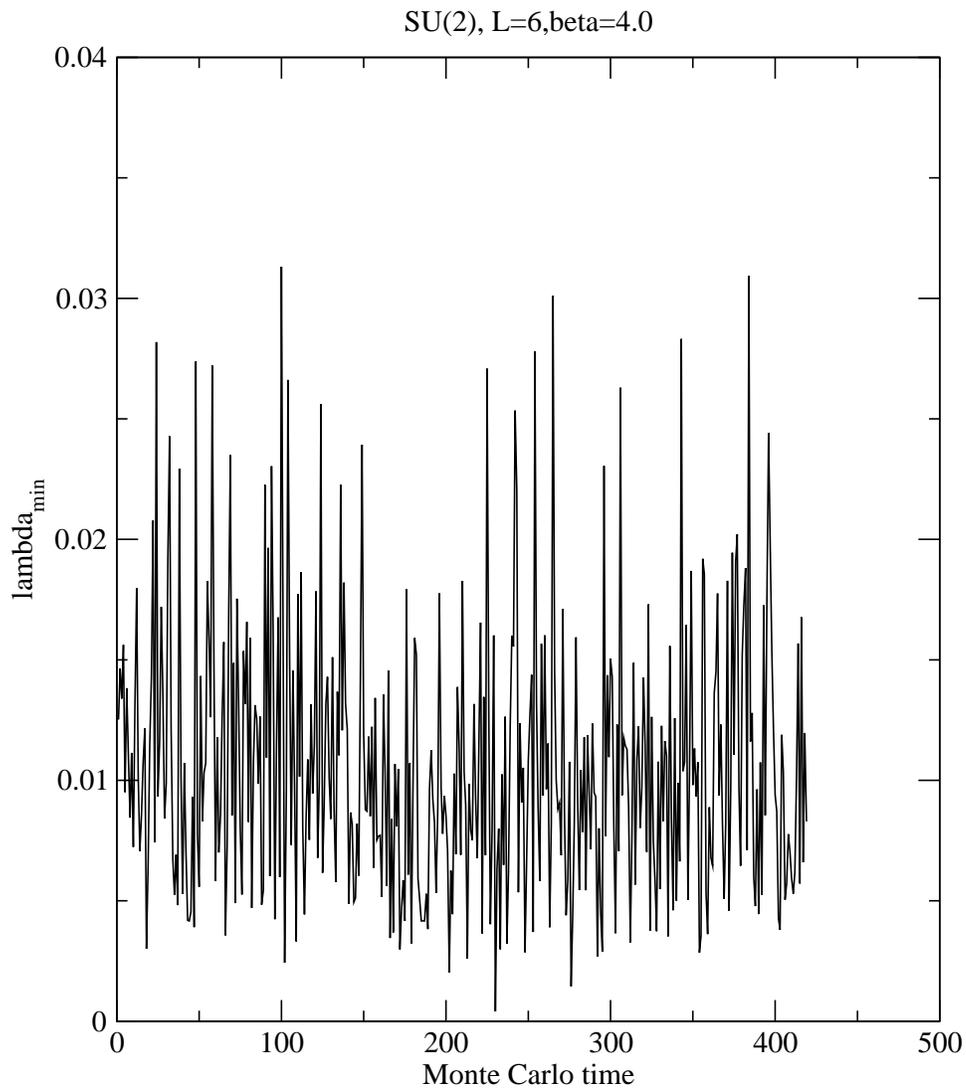
{Minimum eigenvalue for $SU(2)$ theory at $L=6$}

\EPSFIGURE{su2string.eps, width=5.0in}
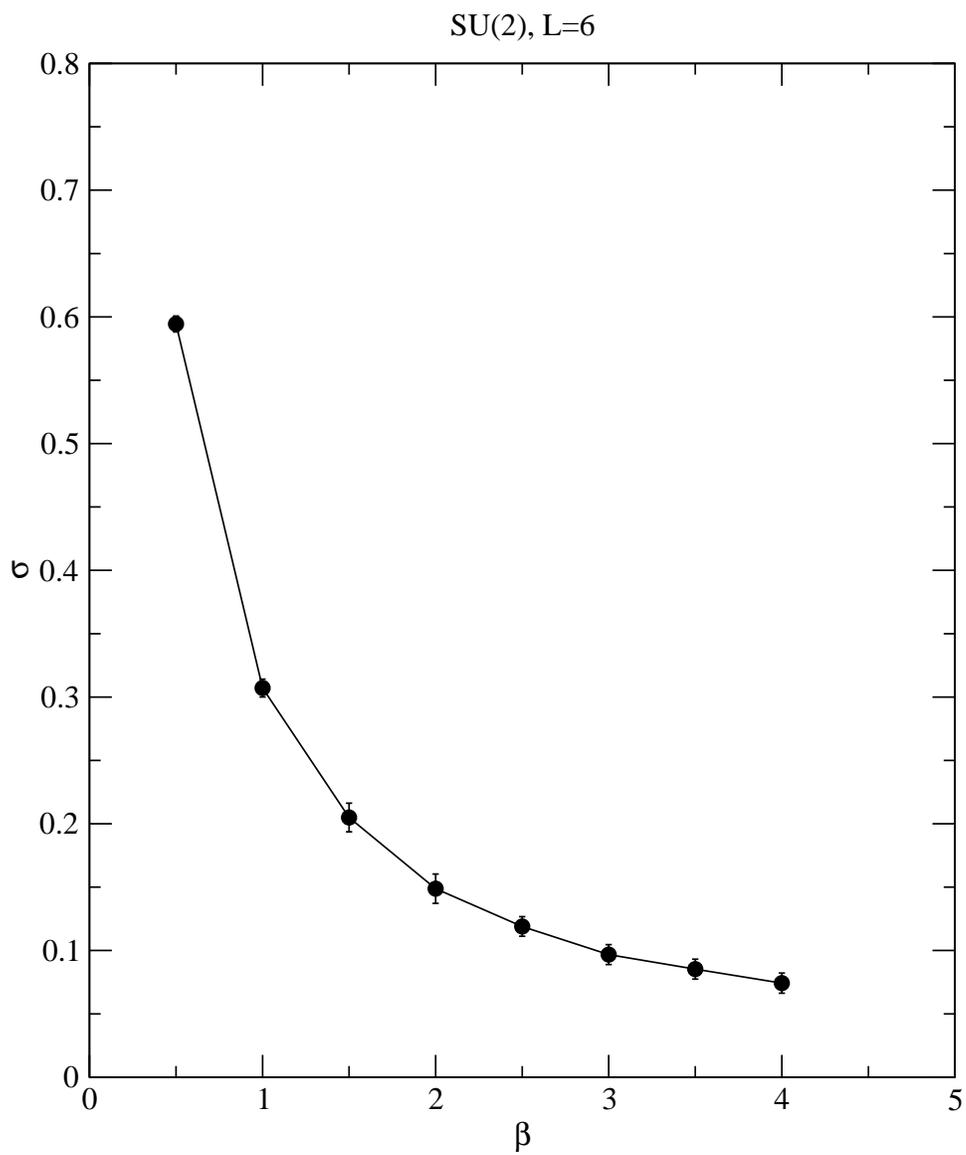
{String tension for $SU(2)$ and $L=6$}

\EPSFIGURE{su2corr.eps, width=5.0in}
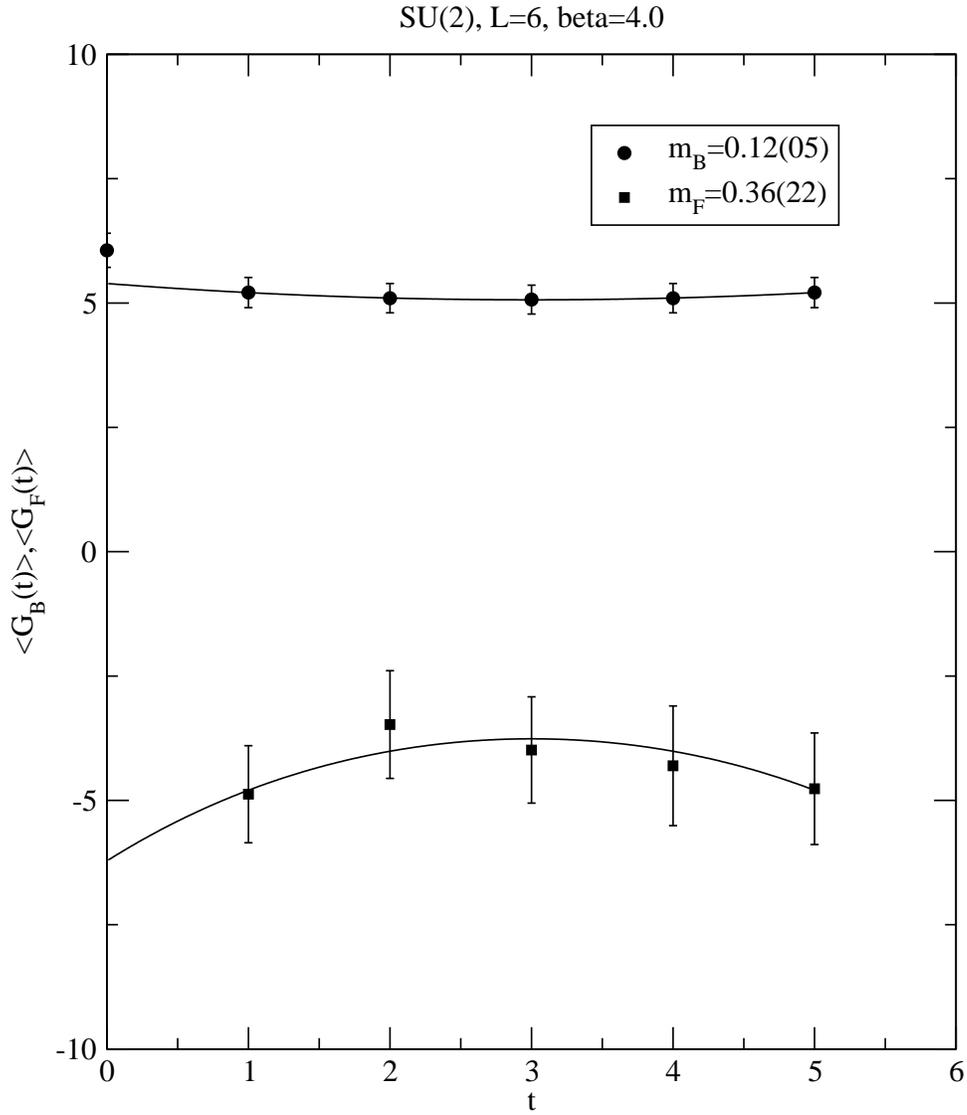
{Bosonic and fermionic correlators in $SU(2)$ for $L=6$ and $\beta=3.0$}

\EPSFIGURE{phasedist.eps, width=5.0in}
{$P(\cos{\alpha})$ vs $\cos{\alpha}$ for $L=3$ $\beta=4.0$}

\EPSFIGURE{phasel6.eps, width=5.0in}
{$<\cos{\alpha}>$ for $L=3$}

\EPSFIGURE{distl3.eps, width=5.0in}
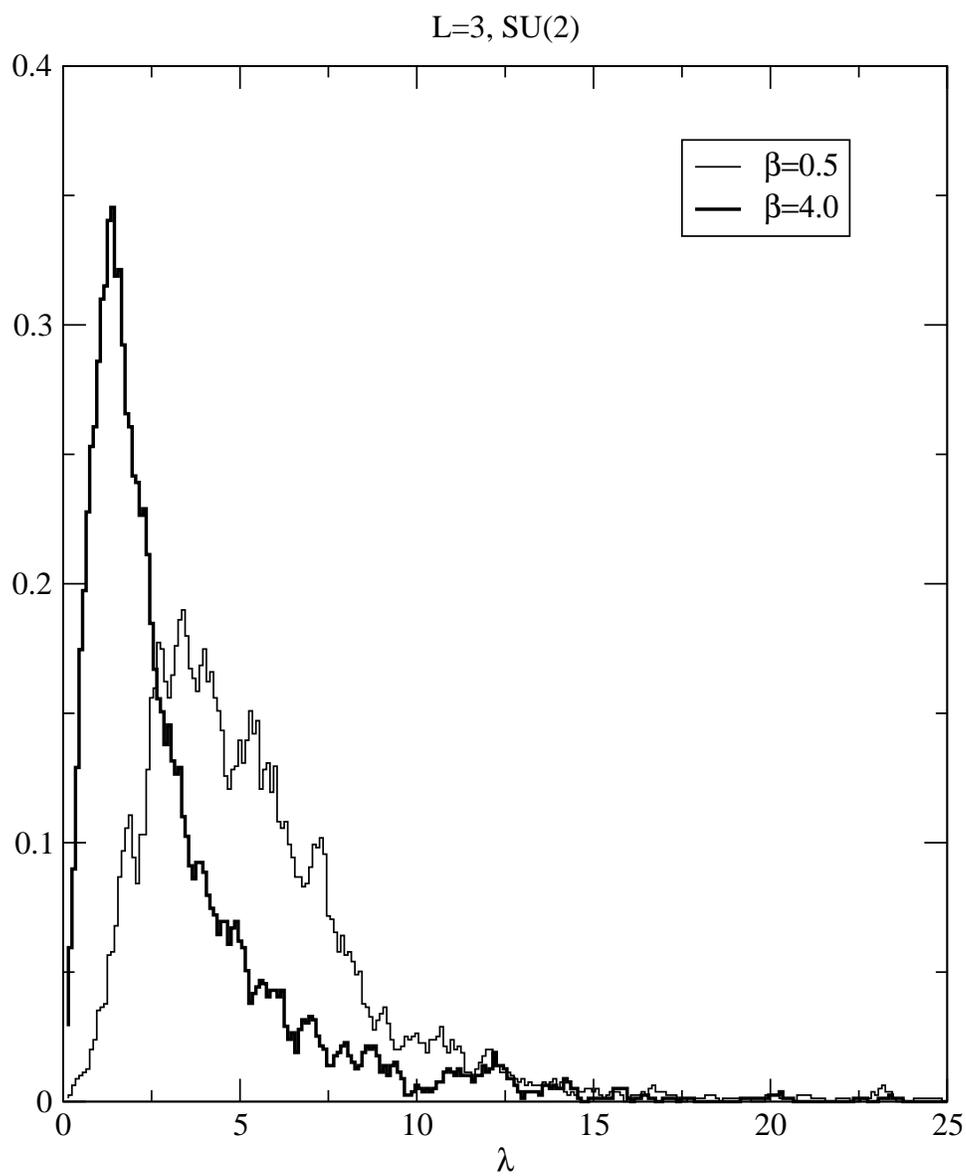
{Distribution of eigenvalues of $\phi_1$ for $SU(2)$, $L=6$}
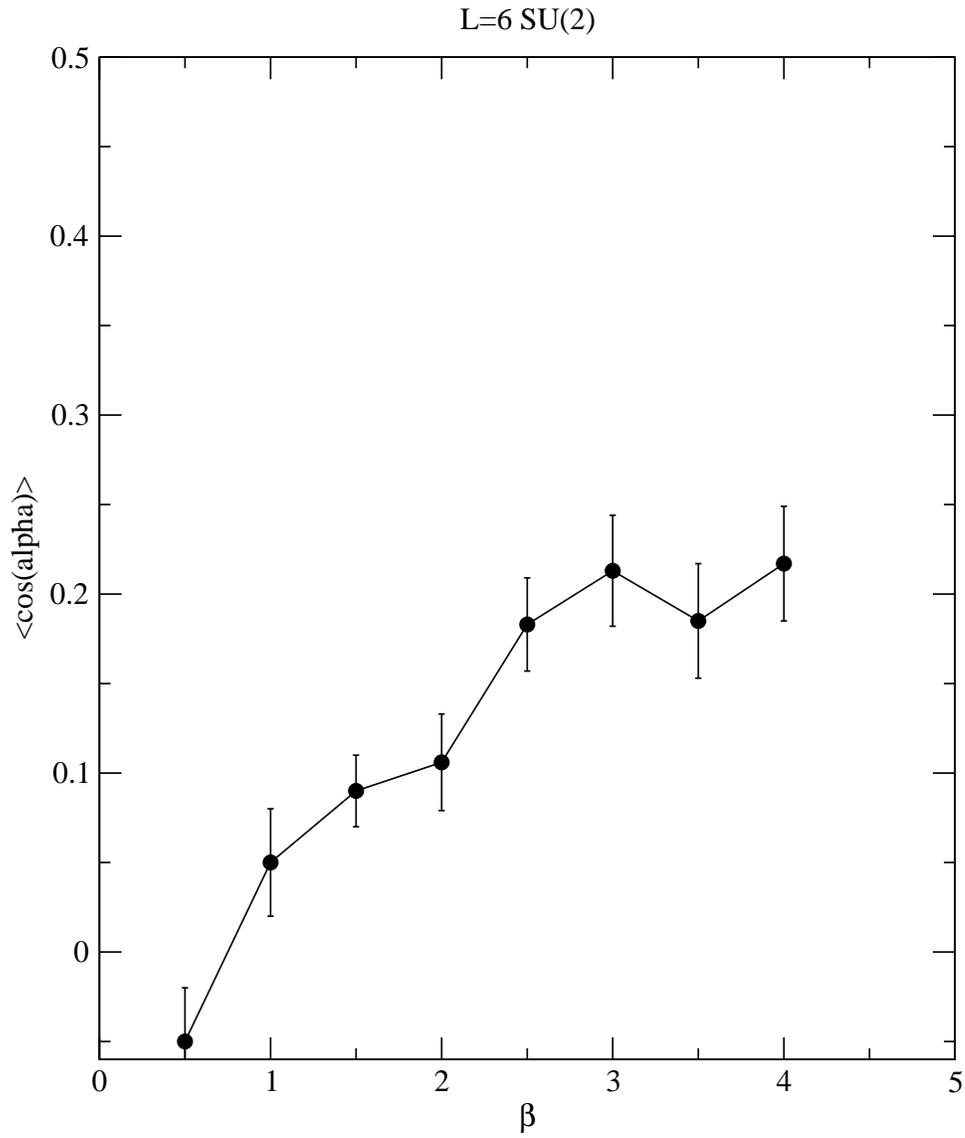
\end{document}